\documentclass{osa-article}

\journal{oe}

\usepackage{amsmath,amssymb}
\usepackage{booktabs}
\usepackage{mathrsfs}
\usepackage{lipsum}
\usepackage{bbm}
\usepackage{graphicx}
\usepackage{caption}
\usepackage{subcaption}
\usepackage{mathtools}
\usepackage{stackrel}
\usepackage{xcolor}
\usepackage{cuted}
\usepackage{flushend}
\usepackage{floatrow}
\DeclareCaptionFont{mycap}{\fontsize{8pt}{11pt}\selectfont #1}
\newcommand{\breathe}[2]{\rule[-#1ex]{0cm}{#2ex}}

\articletype{Research Article}

\begin{document}

\title{On the interplay between physical and content priors in deep learning for computational imaging}

\author{Mo Deng,\authormark{1,5, $\ast$} Shuai Li,\authormark{2,5} Iksung Kang,\authormark{1} Nicholas X. Fang,\authormark{3} and George Barbastathis\authormark{3,4}}

\address{\authormark{1}Department of Electrical Engineering and Computer Science, Massachusetts Institute of Technology, 77 Massachusetts Ave, Cambridge, MA, 02139, USA\\
\authormark{2}SenseBrain Technology Limited LLC, 2550 N 1st Street, Suite 300, San Jose, CA, 95131, USA\\
\authormark{3}Department of Mechanical Engineering, Massachusetts Institute of Technology, 77 Massachusetts Ave, Cambridge, MA, 02139,USA\\
\authormark{4}Singapore-MIT Alliance for Research and Technology (SMART) Centre, One Create Way, Singapore 117543, Singapore\\
\authormark{5}Equal contribution}
\email{\authormark{$\ast$} modeng@mit.edu} 



\begin{abstract}
Deep learning (DL) has been applied extensively in many computational imaging problems, often leading to superior performance over traditional iterative approaches. However, two important questions remain largely unanswered: first, how well can the trained neural network generalize to objects very different from the ones in training? This is particularly important in practice, since large-scale annotated examples similar to those of interest are often not available during training. Second, has the trained neural network learnt the underlying (inverse) physics model, or has it merely done something trivial, such as memorizing the examples or point-wise pattern matching? This pertains to the interpretability of machine-learning based algorithms. In this work, we use the Phase Extraction Neural Network (PhENN) [Optica {\bf4}, 1117-1125 (2017)], a deep neural network (DNN) for quantitative phase retrieval in a lensless phase imaging system as the standard platform and show that the two questions are related and share a common crux: the choice of the training examples. Moreover, we connect the strength of the regularization effect imposed by a training set to the training process with the Shannon entropy of images in the dataset. That is, the higher the entropy of the training images, the weaker the regularization effect can be imposed. We also discover that weaker regularization effect leads to better learning of the underlying propagation model, \textit{i.e.} the weak object transfer function, applicable for weakly scattering objects under the weak object approximation. Finally, simulation and experimental results show that better cross-domain generalization performance can be achieved if DNN is trained on a higher-entropy database, \text{e.g.} the ImageNet, than if the same DNN is trained on a lower-entropy database, \textit{e.g.} MNIST, as the former allows the underlying physics model be learned better than the latter. 
\end{abstract}

\section{Introduction} \label{sec:intro}
\subsection{Two unanswered fundamental questions of deep learning in computational imaging}
Deep learning (DL) has been proven versatile and efficient in solving many computational inverse problems, including image super-resolution\cite{inv:dong14-super-res,inv:perceptual-loss,inv:ledig17,inv:rivernson17-dlm,inv:wang18-super-fluo,inv:nehme18-ML-STORM,deng2018learning}, phase retrieval \cite{inv:sinha17-PhENN,inv:goy2018low, inv:WangH2018,inv:pitkaaho17, inv:ozcan-dnn-extDOF, ren2018learning,deng2020learning,inv:deng2020probing}, imaging through the scattering medium \cite{inv:horisaki16,inv:IDiffNet,li2018deep}, optical tomography \cite{kamilov2015learning,kamilov2016optical,inv:goy19-3Dtomo} and so on. See \cite{inv:barbastathis19-review,mccann2017convolutional} for more detailed reviews. Besides the superiority of performance over classical approaches in many cases, the DNN-based methods enjoy the advantage of extremely fast inference after the completion of the training stage. It is during the latter that the DNN weights are optimized as the training loss function between ground truth objects and estimated objects is reduced.\\

\noindent Despite great successes, two important questions remain largely unanswered: first, how well does a model trained on one dataset generalize directly to objects in disjoint classes; second, how well does the trained neural network learn the underlying physics model? These questions are well-motivated, since access to a large number of training data in the same category with those in the test set is not always possible and it would reduce the practicality of deep learning if the model trained on one set cannot reasonably well generalize to the other. Moreover, one major skepticism against deep learning is: has the algorithm actually learnt anything about the underlying physics or is it merely doing some trivial point-wise denoising, pattern matching, or, worse, just memorizing and reproducing examples from the training set? \\

\noindent In this paper, we recognize that these two questions are directly related: if the trained DNN were able to perfectly learn the underlying (inverse) physics law, which is satisfied unconditionally by all classes of objects, the prediction performance would not have degraded even if tested on very different examples. Conversely, if the DNN does not learn the model well and instead relies on regularization, i.e. priors on the examples, to reconstruct, then cross-domain generalization will be problematic.  \\

\noindent Though training datasets typically contain at least hundreds, if not thousands, training examples, the DNNs are typically highly over-parameterized models; that is, the number of trainable parameters is larger compared to the number of the training examples. Such under-determined, nonlinear system necessitates measures be taken to ensure the learned DNN to correspond to the true physics model. Our investigation unearths one such important factor, the choice of training examples, using the Phase Extraction Neural Network (PhENN) \cite{inv:sinha17-PhENN} for lensless phase retrieval as an example. In fact, we discover that a trained DNN corresponds to the physics law better, and therefore cross-domain generalizes better, when the training set is more generic, e.g. the ImageNet \cite{deng2009imagenet}, than if it is more constrained, e.g. the MNIST \cite{nn:lecun2010mnist}. Therefore, when encountering insufficient training data in the same class as the test data, which is very common, the best compromise is to train the neural network on a less constrained publicly available standardized dataset such as the ImageNet, with reasonable confidence that it produces the reconstructions accurately in the domain of interest. 

\subsection{Phase retrieval and the weak object transfer function (WOTF) for lensless phase imaging} \label{subsec:wotf-general} 

\begin{figure}[ht!]
\centering\includegraphics[width=\linewidth]{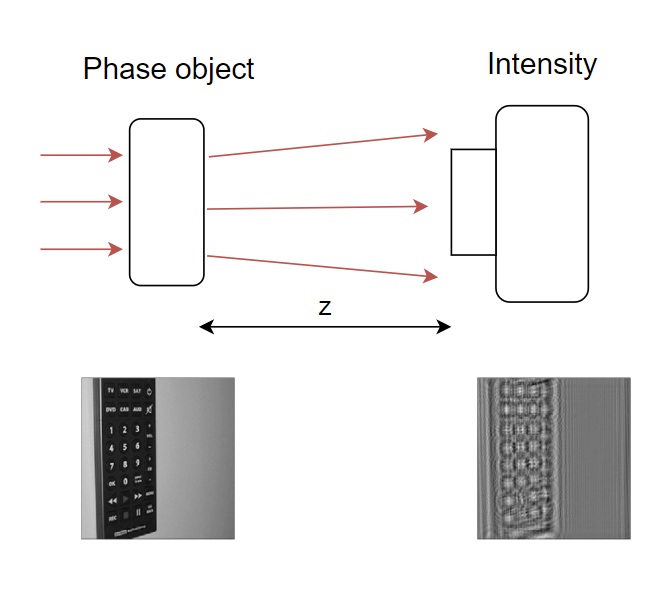}
\caption{Schematic plot of the lensless phase imaging system.}
\label{fig:LPI}
\end{figure}

In the lensless phase imaging system (Fig.~\ref{fig:LPI}), the phase object is illuminated by collimated monochromatic light. The light transmitted through the phase object propagates in free space and forms an intensity pattern on the detector placed at a distance $z$ away. Assuming that the illumination plane wave has unit amplitude, the forward model can be described as
\begin{equation}
g(x,y)=\left\vert\text{exp}\left\{\rule[-1ex]{0cm}{3ex}if(x,y)\right\}*\text{exp}\left\{\rule[-1ex]{0cm}{3ex}i\frac{\pi}{\lambda z}\left(x^2+y^2\right)\right\}\right\vert^{2}.
\label{eq:forward}
\end{equation}
Here, $f(x,y)$ is the phase distribution of the object, $g(x,y)$ is the intensity image captured by the detector, $\lambda$ is the wavelength of the illumination light, $i$ is the imaginary unit and $*$ denotes the convolution operation.  \\

\noindent Eq.(\ref{eq:forward}) is non-linear. However, when the weak object approximation holds, $\exp\{if(x,y)\}\approx 1+if(x,y)$ \cite{tian2015quantitative}, the forward imaging model may be linearized as
\begin{equation}
G(u,v)\approx\delta(u,v)+2\sin\left(\pi\lambda z(u^2+v^2)\right)F(u,v).
\label{eq:forward-weak}
\end{equation}
Here, $G(u,v)$ and $F(u,v)$ are the Fourier transforms of the intensity measurement $g(x,y)$ and the phase distribution of the object $f(x,y)$, respectively; $\sin\left(\pi\lambda z(u^2+v^2)\right)$ is the weak object transfer function (WOTF) for lensless phase imaging. The nulls of the WOTF are of particular significance as sign transitions surrounding these nulls in the frequency domain would cause a $\pi$ phase shift in the measurement. We will address further on this effect in Section \ref{subsec:wotf} and \ref{subsec:star}. 

\subsection{Phase Extraction Neural Networks (PhENN)} \label{subsec:phenn-intro}
The Phase Extraction Neural Network (PhENN) \cite{inv:sinha17-PhENN} is a deep learning architecture that can be trained to recover an unknown phase object from the raw intensity measurement obtained through a lensless phase imaging system. Since PhENN was proposed, three general types of strategies have been followed to further enhance its performance. The first category focused on optimizing the network architecture or training specifics of PhENN, including the network depth, the training loss functions etc. \cite{li2019analysis} The second category focused on optimizing the spatial frequency components of the training data to compensate imbalanced fidelity in the high and low frequency bands in the reconstructions. The same rationale applies not only to phase retrieval but many other applications as well. Li \textit{et al.} proposed a spectral pre-modulation approach \cite{inv:PhENN-spectral-premod} to amplify the high frequency content in the training data and experimentally demonstrated that the PhENN trained in this fashion achieved a better spatial resolution while at the cost of the overall reconstruction quality. Subsequently, Deng \textit{et al.} proposed a learning synthesis by deep neural network (LS-DNN) method \cite{deng2018learning}, which is able to achieve a full-band and high quality reconstruction in phase retrieval and other computational imaging applications, by splitting, separately processing and recombining the low and high spatial frequencies. The third category is to make the learning scheme more physics-informed. Attempts are made where, unlike the original PhENN, the forward model is incorporated via model-based pre-processing \cite{inv:goy2018low,deng2020learning,inv:deng2020probing}, etc. and such strategy is proven particularly useful under most ill-posed conditions. However, since all these efforts are secondary to the main objectives of this paper, we choose not to implement them.\\

\begin{figure}[ht!]
    \centering
    \includegraphics[width=\textwidth]{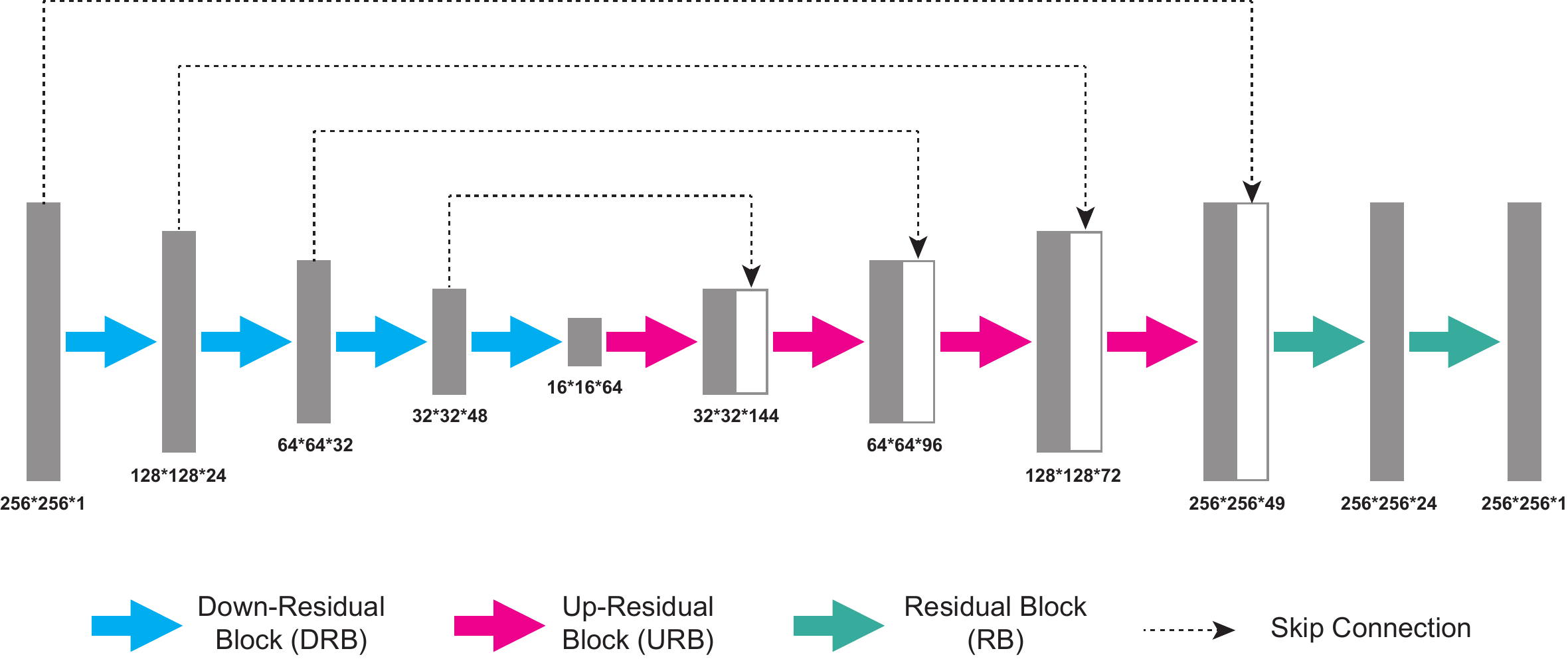}
    \caption{The general architecture of PhENN.}
    \label{fig:PhENN-generall}
\end{figure}

\noindent The PhENN architecture used in this paper is shown in Fig.~\ref{fig:PhENN-generall}. It uses an \textit{encoder-decoder} structure with skip connections, a structure proven efficient and versatile in many applications of interest. The \textit{encoder} consists of 4 Down-Residual blocks (DRBs) to gradually extract compressed representations of the input signals to preserve high-level features; subsequently, the \textit{decoder} component comprises of 4 Up-Residual blocks (URBs) and two (constant-size) Residual blocks (RBs) to expand the size of the feature maps and form the final reconstruction. To best preserve the high spatial frequencies, skip connections are used to bypass the feature maps from the \textit{encoder} to the corresponding layers of the same size in the \textit{decoder}. More details about the architecture of each functional blocks are available in Appendix A. Compared to the original PhENN \cite{inv:sinha17-PhENN}, in this paper we implemented two modifications: first, we started from intensity patterns of size $256 \times 256$, same as that of the objects, as only with the same pixel patch sizes and number of pixels, can the computation of the weak object transfer function (WOTF) (see Section. \ref{subsec:wotf-general}) be physically sound. Second, PhENN is trained with negative Pearson correlation coefficient (NPCC), a loss proven to be more beneficial for restoring fine details of the objects \cite{inv:IDiffNet}, instead of using the mean absolute error (MAE), a pixel-wise loss known to suffer from the oversmoothing issue. The exact form of NPCC is given later in Eq. (\ref{eq:npcc}).\\

\subsection{Generalization error in machine learning} \label{subsec:previous}

\noindent Previous works have addressed different aspects of the generalization errors of DNNs. Some \cite{neyshabur2017exploring,neyshabur2018towards} aimed at tightening bounds on the capacity of a model, \textit{i.e.} the number of training examples necessary to guarantee generalization. Ref. \cite{neyshabur2017pac} provided a finer generalization bound for a subclass of neural networks in terms of the product of the spectral norm of their layers and the Frobenius norm of their weights. Yet, beautiful as the works are, the model capacity bounds and the generalization error bounds are not tight enough to provide practical insights. In \cite{zhang2016understanding}, the authors provide insights into the explicit and implicit role of various regularization methods (not to confuse with the regularization effect imposed by the training data, as mentioned in this paper) to reduce generalization error, including using dropout, data augmentation, weight decay, \textit{etc}, practices that are commonly applied in the deep learning implementation, including in PhENN. Other works examined the generalization dynamics of large DNNs using SGD \cite{advani2017high}, the generalization ability of DNNs with respect to their robustness \cite{xu2012robustness}, \textit{i.e.} the ability of the DNN to cope with small perturbations of its input. As \cite{jakubovitz2019generalization}, a nice review on this general topic, points out, one of the open problems in this area is to understand the interplay between memorization and generalization. This work, unlike in previous works where generalization typically refers to generalization to unseen examples similar to the training dataset, attempts to provide some insights into how the choice of training dataset might affect the cross-domain generalization performance, by an empirical investigation. \\

\section{Entropy as a metric of the strength of the regularization effect imposed by a training set}

\noindent When a dataset is selected as the training dataset, its influence to the training is reflected by the regularization effect it imposes. For example, when examples with similar statistical distributions are presented to the DNN as the ground truth, the training, or the optimization of the weights in the DNN is susceptible to taking the shortcut of ``memorizing" the examples so as to minimize the training loss, instead of learning the underlying physics law, making such trained DNN produce reconstructions similar to those training examples. The cross-domain generalization therefore becomes problematic. Under such scenarios, the overly constrained prior information represented by the training set imposes an unduly strong regularization effect to the training. \\

\noindent Intuitively, ImageNet \cite{deng2009imagenet} contains natural images of a broad collection of contents and scenes, and thus should present a more generic prior. On the other hand, MNIST \cite{nn:lecun2010mnist}, a database with the much more restricted set of handwritten digits only, would be expected to impose stronger regularization. However, to quantify the regularization effect imposed by a particular training dataset is not as straightforward. Here, we employ Shannon entropy \cite{shannon1948mathematical,cover2012elements} of the images in the dataset for that purpose: the higher the entropy, the weaker the regularization effect this dataset imposes to the training process. \\

\noindent Shannon entropy is a measure of the uncertainty of a random variable. Let $X$ denote the random variable with the distribution $p(X)$. Without loss of generality, we assume that $X$ is discrete and assumes a finite alphabet $\mathscr{X}=\{x_1,\cdots,x_K\}$, then the entropy (in bits) of $X$, under the distribution $p(\cdot)$, $H_p(X)$, is defined as:
\begin{equation}
H_p(X)=-\sum_{k=1}^{K}p(x_k)\text{log}_{2}p(x_k),
\label{eqn:entropy-discrete}
\end{equation}
where $p(x_k):=\text{Pr}\{X=x_k\}$. The well-known Asymptotic Equipartition Theorem \cite{shannon1948mathematical,cover2012elements} states that, with the obvious constraint $\sum_{k}p(x_k)=1$, the entropy is maximized at the equiprobable case, \textit{i.e.} $p(x_k)=\frac{1}{K}$ for all $i$'s and the maximized entropy is $\text{log}_{2}K=\text{log}_{2}\lvert \mathscr{X}\rvert$. The higher the entropy, the higher the level of uncertainty is in the source. Conversely, if the distribution is completely deterministic (\textit{i.e.} $p(x_i)=1$ for some $i$ and $p(x_j)=0$ for $j\neq i$), the entropy of the source is 0, the lowest extreme. \\

\noindent For an image $f$ of size $M\times N$, defined on the alphabet $\mathscr{X}^{M\times N}$, the empirical distribution on the pixel value $X \in \mathscr{X}$, is defined as 
\begin{equation}
\hat{p}(x_k)=\frac{\sum_{i,j}\mathbbm{1}\{f_{i,j}=x_k\}}{M \times N}=\frac{\text{number of pixels of value $x_k$}}{M\times N},
\label{eqn:empirical-dist}
\end{equation}
and the entropy of this image $f$ can be approximated by the entropy $H_{\hat{p}}(\cdot)$ according to this empirical distribution $\hat{p}(\cdot)$, according to Eq.(\ref{eqn:entropy-discrete}). \\

\begin{figure}[hbt!]
    \centering
    \includegraphics[width=0.9\textwidth]{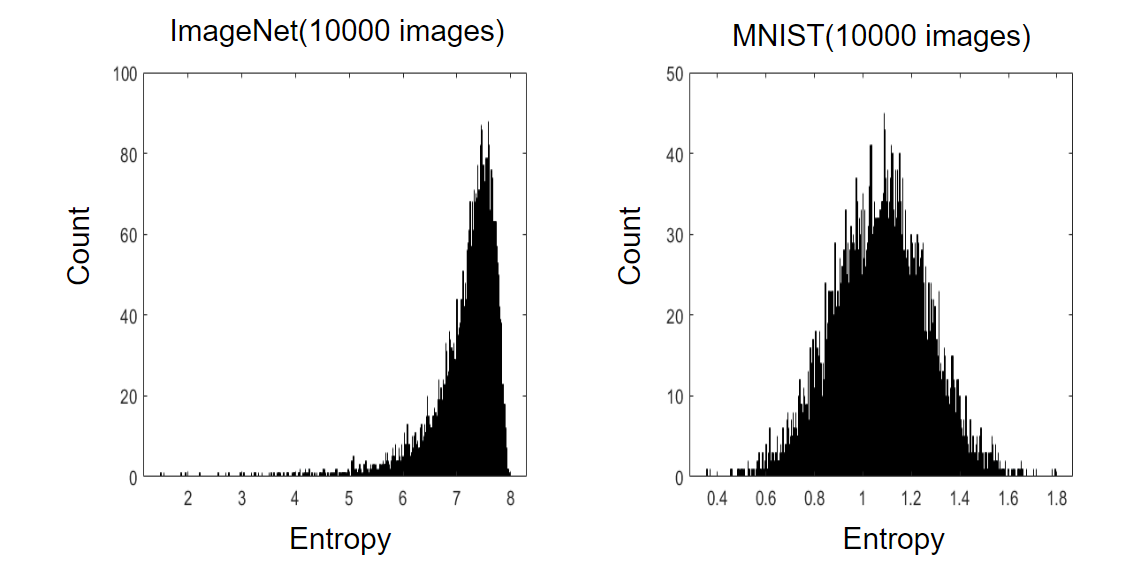}
    \caption{Entropy histogram of ImageNet and MNIST. Each histogram is computed based on 1000 bins and 10000 images.}
    \label{fig:entropy-comparison}
\end{figure}

\noindent In Fig.~\ref{fig:entropy-comparison}, we showed a histogram of our computed Shannon entropy of the images in ImageNet and MNIST, two representative standardized databases. 10000 8-bit images were selected from each set and a histogram of the entropy of the images is computed based on 1000 bins between 0 to 8 bits. The ImageNet images generally have higher entropy (mean 7.190 and standard deviation 0.600) and most images have their entropy between 7 and 8 bits; while the MNIST images have low entropy (mean 1.072 and standard deviation 0.185). Interestingly, this matches well with our anticipation that the MNIST images are approximately binary images -- the entropy of the image will be 1 in the case of a perfect binary image with equal densities, \textit{i.e.} $p(x_i)=p(x_j)=0.5, \text{for } i\neq j$ and $p_k=0$ for all other $k$'s. The deviation of the observed entropy in MNIST images from the ideal entropy of perfect binary images comes from the fact that the nonzero densities are not equal in the empirical distribution of MNIST images. (and later we will see that such narrow distribution of Shannon entropy found in MNIST makes the reconstruction of WOTF challenging as it offers very limited insight of out-of-focus images).\\

\noindent For a database, the connection between the entropy of its images and the strength of the regularization effect can be argued as follows: the higher the entropy of images in a database have, the closer the empirical distributions are to the uniform distribution, which is the extreme case where the weakest regularization effect can be imposed. Consider the extreme case where pixels on an image admit an i.i.d. uniform distribution on $\mathscr{X}$, the empirical distribution in Eq. (\ref{eqn:empirical-dist}) is a uniform one. In this extreme case, no regularization effect can possibly be imposed by the training examples since if the DNN were to ``memorize" such training examples, nothing but the totally random distribution on pixel-values could be remembered. On the other extreme, if the training dataset contains zero-entropy images only (all pixels are identical on each image), the training loss can be minimized merely by the DNN learning to produce such uniform-valued images, totally neglecting the underlying physics model associated with the stochastic processes involved during the image formation. In the next section we show that the stark difference in entropy between ImageNet and MNIST shown in Fig.~\ref{fig:entropy-comparison} strongly implies respective difference between the two datasets in terms of generalization ability.\\

\section{Results on synthetic data}
\subsection{Performance of cross-domain generalization under different training datasets}

In this section, we compare the cross-domain generalization performance of PhENN to ImageNet, MNIST and two other datasets, the IC layout dataset \cite{inv:goy2018low} and Face-LFW, when trained on ImageNet and MNIST, respectively. The choice of IC layout database and Face-LFW is well-motivated as they are both rather distinct from each training database; yet, the former is somewhat similar to the MNIST dataset in the piecewise constancy but the latter does not have this piece-wise constancy. \\

\noindent The intensity measurement is synthetically generated by the standard phase retrieval optical apparatus in Fig.~\ref{fig:LPI}, according to Eq.(\ref{eq:forward}). All training and testing objects are of size $256\times256$, with the pixel size of the objects and on the detector being $20 \mu\text{m}$. The propagation distance is set to be $ z=100 \text{mm}$. In order to qualify the weak object approximation, so that the learned WOTF can be explicitly computed from the measurements and the corresponding phase objects, the maximum phase depth of the objects is kept below $0.1 \pi$ rad. In simulation, the weak objects are obtained from applying a heuristic one-to-one calibration curve, say a linear one, that maps from the 8 bit alphabet $\{0, 1,\cdots, 255\}$ to one that lies within 0 and $0.1\pi$ rad, as the entropy of an image is invariant under one-to-one mappings and therefore, we can have the weak object approximation and the entropy distributions above hold concurrently. \\

\noindent During the training stage, pairs of intensity measurements and weak phase objects are used as the input and output of PhENN, respectively. The disparity between the estimate produced by the current weights in PhENN with the ground truth is used to update the weights using backpropagation. After PhENN is trained, in the test stage, the measurements corresponding to unseen examples are fed into the trained PhENN to predict the estimated phase objects. The loss function used in training is the negative Pearson correlation coefficient (NPCC), which has been proved helpful for better image quality in reconstructions \cite{inv:IDiffNet,inv:goy2018low,deng2020learning,inv:goy19-3Dtomo, li2019analysis}. For phase object $f$ and the corresponding estimate $\hat{f}$, the NPCC of $f$ and $\hat{f}$ is defined as
\begin{equation}
\text{NPCC}(\hat{f},f)\equiv
\frac {\displaystyle{-\sum_{x,y}}\left(\breathe{1}{3} \hat{f}(x,y)-\left<\hat{f}\right>\right)\:
\left(\breathe{1}{3} f(x,y)-\left<f\right>\right)}
{\sqrt{\displaystyle{\sum_{x,y}}\left(\breathe{1}{3} \hat{f}(x,y)-\left<\hat{f}\right>\right)^2}\:
\sqrt{\displaystyle{\sum_{x,y}}\left(\breathe{1}{3} f(x,y)-\left<f\right>\right)^2}},
\label{eq:npcc}
\end{equation}
where $\left<\hat{f}\right>$ and $\left<f\right>$ are the spatial averages of the reconstruction $\hat{f}$ and true object $f$, respectively. If the reconstruction is perfect, we have $\text{NPCC}(\hat{f},f)=-1$. One caveat of using NPCC is that a good NPCC metric cannot guarantee the reconstruction is of the correct quantitative scale, as NPCC is invariant under affine transformations, \textit{i.e.}, $\text{NPCC}(\hat{f},f)=\text{NPCC}(a\hat{f}+b,f)$ for scalars $a$ and $b$. Therefore, to correct the quantitative scale of the reconstructions without altering the PCC value, a linear fitting step is carried out based on the validation set and the learned $a,b$ values are used to correct the test sets. The quantitative performance of the reconstructions on various datasets by ImageNet-trained PhENN and MNIST-trained PhENN, are compared in Table~\ref{table:quantitative-cross-generalization}. The quantitative metrics chosen considered both the pixel-wise accuracy through MAE and the structural similarity through PCC.\\

\begin{table*}[hbt!]
\begin{center}
\begin{tabular}{|l||c|c||c|c|}
 \hline
& \multicolumn{2}{c||}{Average PCC $\pm$ std.dev}  & \multicolumn{2}{c|}{Average MAE $\pm$ std.dev} \\
 \hline
  & Train on ImageNet & Train on MNIST & Train on ImageNet & Train on MNIST \\
 \hline
Test on ImageNet & 0.932 $\pm$ 0.068 & 0.578 $\pm$ 0.213 & 0.032 $\pm$ 0.024 & 0.055 $\pm$ 0.021 \\
 \hline
Test on MNIST  & 0.964 $\pm$ 0.039 & 0.9998 $\pm$ 0.0002 & 0.006 $\pm$ 0.001 & 0.0008 $\pm$ 0.0004 \\
 \hline 
Test on IC & 0.942 $\pm$ 0.014 & 0.866 $\pm$ 0.051 & 0.018 $\pm$ 0.004 & 0.024 $\pm$ 0.006 \\
\hline 
Test on Face-LFW & 0.969 $\pm$ 0.015 & 0.758 $\pm$ 0.104 & 0.016 $\pm$ 0.004 & 0.024 $\pm$ 0.008\\
\hline 
\end{tabular}
\caption{\textbf{Cross-domain generalization ability performance of PhENN trained with ImageNet and MNIST, respectively, for $ z=100 \text{ mm}$ (synthetic data).}}
\label{table:quantitative-cross-generalization}
\end{center}
\end{table*}

\noindent From Table~\ref{table:quantitative-cross-generalization}, we make the following interesting observations: 
\begin{itemize}
    \item It is always ideal to train with examples that are in the same class with the test examples.In both metrics and for tests on both ImageNet and MNIST, same-domain generalization outperforms cross-domain generalization. Among same-domain generalization, more constrained prior information (training dataset with lower entropy), and thus the stronger prior information, is beneficial, since it more strongly regulates the reconstructions to be similar to those examples from training, and fortunately, also those from the test.
    
    \item When access to training data in the same database as the test data is not possible, if the model is trained on ImageNet, the cross-domain generalization to more constrained MNIST is satisfactory; however, such performance is catastrophic in the opposite case, when the model is trained on more constrained MNIST but tested on the more generic ImageNet. 
    
    \item Likewise, when tested on the IC layout dataset and the Face-LFW dataset, the cross-domain generalization is always better if PhENN is trained on ImageNet than on MNIST, despite the fact that the IC layout dataset shares with MNIST, but not with the ImageNet, the feature of piece-wise constancy. However, such performance gap seems to depend on the (dis)similarity between the test datasets and the training set (ImageNet or MNIST) --- the performance gap (cross-domain generalization performance difference between MNIST-trained PhENN and ImageNet-trained PhENN) is much larger when the test set is Face-LFW than IC.  
\end{itemize}

\begin{figure*}[hbt!]
    \centering
    \captionsetup{justification=centering}
    \includegraphics[width=\textwidth]{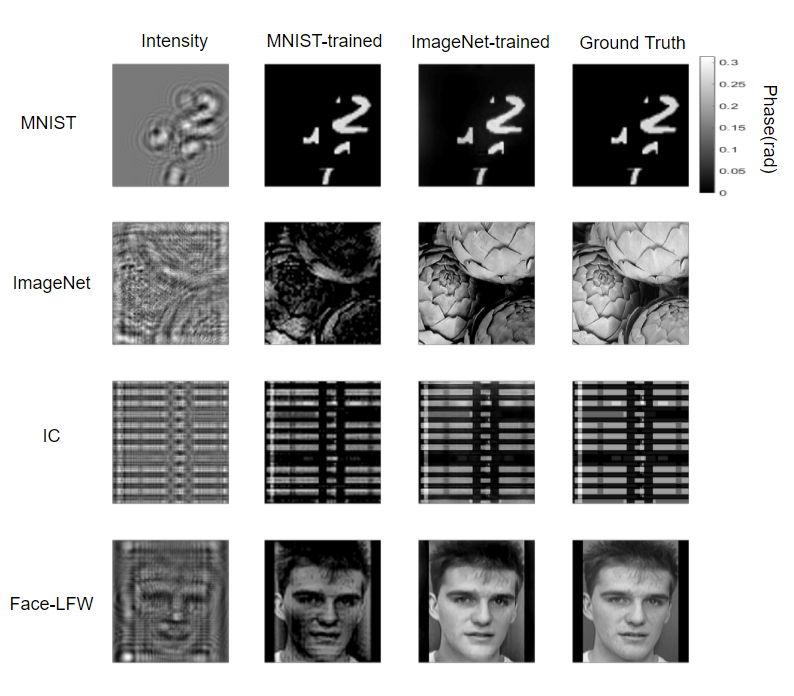}
    \caption{Cross-domain generalization performances of ImageNet-PhENN and MNIST-PhENN on synthetic data. The defocus distance is  $ z=100\text{mm}$.}
    \label{fig:cross-generalization}
\end{figure*}

\noindent More interesting observations are available in Fig.~\ref{fig:cross-generalization}, where we show representative examples of predicting objects in various classes when PhENN is trained by ImageNet and MNIST, respectively. From these examples, the regularization effect imposed by the MNIST training dataset is clear: PhENN has been forced to inherit the piece-wise constant and sparse features from the MNIST examples and pass them to the reconstructed ImageNet, Face-LFW and IC layout data, which caused distortions of different extent. It is also noteworthy that the IC objects share with the MNIST objects on the piecewise constant features (though differ from the latter on the sparsity priors), making the degradation caused by the regularization effect imposed by MNIST dataset to reconstructed ICs much less severe than that to ImageNet and Face-LFW objects. 

\subsection{How well has PhENN learned the physics model? } \label{subsec:wotf}
To quantitatively verify that ImageNet-trained PhENN learns the underlying physics law better than MNIST-trained PhENN, we compare the \textit{learned} WOTF (LWOTF) of the ImageNet-trained PhENN and MNIST-trained PhENN.\\

\noindent Once the network has been trained, based on a test set of $K$ test images, the LWOTF is computed as,
\begin{equation}
\text{LWOTF}=\frac{1}{K}\sum_{k=1}^{K}\frac{G_{k}(u,v)-\delta(u,v)}{\hat{F}_{k}(u,v)},
\end{equation}
where $G_{k}(u,v)$ and $\hat{F}_{k}(u,v)$ are the Fourier transforms of the intensity measurement $g_{k}(x,y)$ and the network's estimated phase $\hat{f}_{k}(x,y)$ for the $k$th testing object, respectively. For better generality, we split the test set of $K=100$ images into four equally large subsets, that is, 25 test images from the ImageNet, MNIST, Face-LFW and IC layouts each. We denote the LWOTF of the ImageNet-trained PhENN and MNIST-trained PhENN as LWOTF-ImageNet and LWOTF-MNIST, respectively. \\

\begin{figure*}[ht!]
\centering\includegraphics[width=\textwidth]{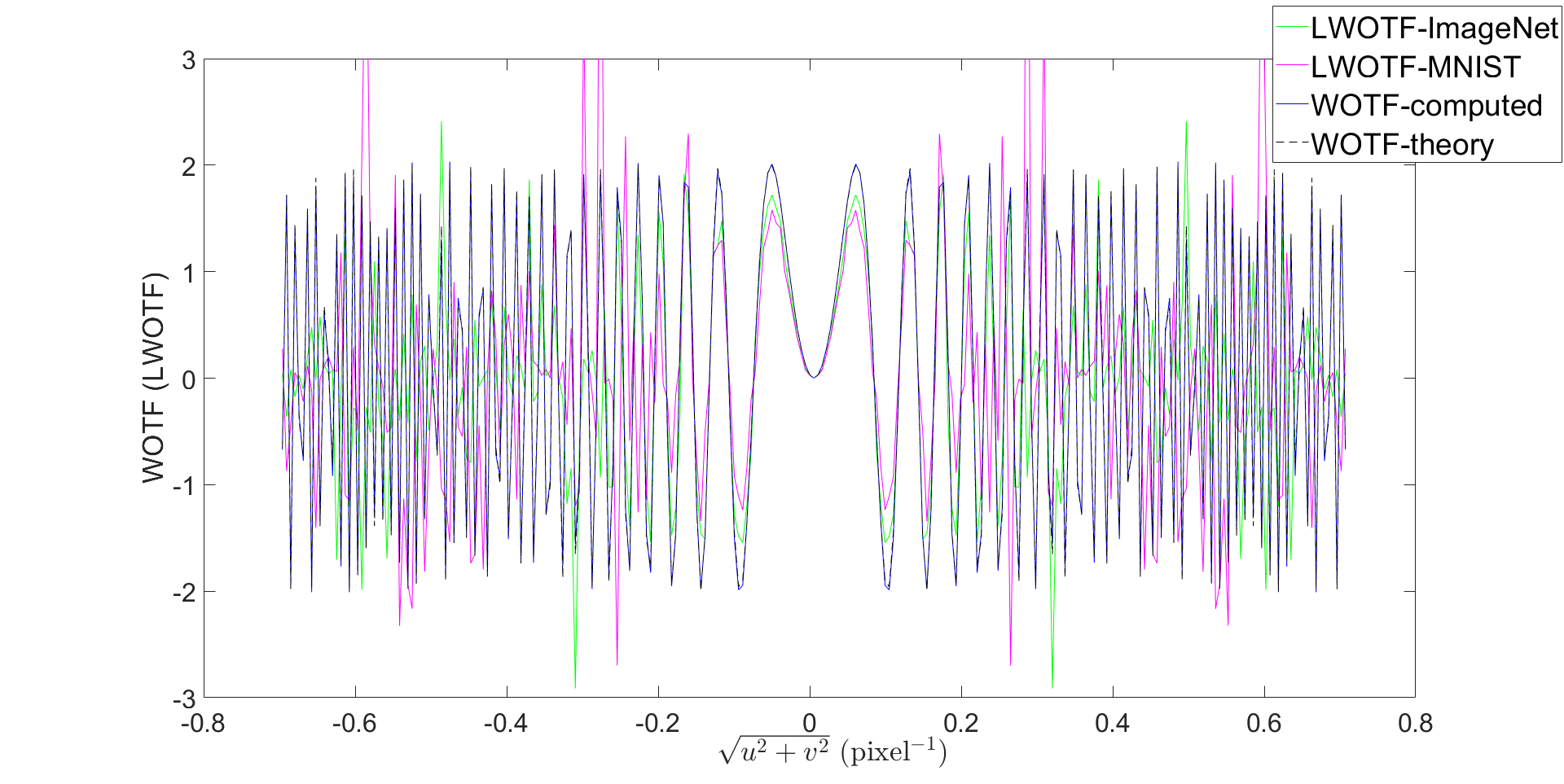}
\caption{Comparison of LWOTFs-ImageNet, LWOTF-MNIST, WOTF-computed and WOTF-theory. The values are cropped to the range of $[-3,3]$ and outliers from LWOTF-MNIST have been cropped out.}
\label{fig:WOTF-comparison-simulation}
\end{figure*}

\noindent In Fig.~\ref{fig:WOTF-comparison-simulation}, we show the 1D cross-sections along the diagonal directions of the WOTF (computed from the ground truth examples), LWOTF-ImageNet and LWOTF-MNIST, respectively. Also, we plot the theoretical WOTF $\text{sin}(2\lambda z(u^2+v^2)$), denoted WOTF-theory, under the same sampling rate as the detection process. WOTF-theory is indistinguishable from WOTF-computed, indicating that the weak object approximation holds well. For better visualization, values are cropped to the range $[-3,3]$ (the values cropped out are outliers, all from LWOTF-MNIST). We find that ImageNet-trained PhENN indeed learned the WOTF better than the MNIST-trained PhENN. Also, we note the mismatch between the LWOTF-ImageNet and the WOTF-theory becomes larger at higher spatial frequencies, which is due to the under-representation of the high spatial frequencies in the reconstructions that has been extensively argued in \cite{inv:PhENN-spectral-premod,deng2020learning}. In this paper, we choose not to overcome this limitation by the LS-DNN technique \cite{deng2018learning,deng2020learning}, so that the choice of training dataset is the only difference between ImageNet-PhENN and MNIST-PhENN. \\

\noindent Besides the direct WOTF comparison, we propose an alternative study where we verify that ImageNet-trained PhENN learns the propagation model better than the MNIST-trained PhENN. From Eq. (\ref{eq:forward-weak}), we see that there exist several nulls (the locations where the values of the transfer function are equal to zero) in the weak object transfer function and at those nulls, the sign of the transfer function switches, introducing a phase delay of $\pi$ rad in the spatial frequency domain. As a result, the measured pattern at the detector plane will shift by half period in the spatial domain at those frequencies. We refer this phenomenon as the ``phase shift effect". Because of that, when we image a star-like binary weak phase object with $P$ periods, the fringes in the measurement will become discontinuous (see Fig.~\ref{fig:star-compare} later). In particular, for a defocus distance $z$ (in mm), the radii of discontinuity $r_k$ for $k=1,2,\cdots$, and the associated spatial frequency in $(u_k, v_k)$ jointly satisfy
\begin{equation}
\lambda z(u_{k}^2+v_{k}^2)=\lambda z (\frac{P}{2\pi r_k})^2= k.
\label{eq:discon-frequency}
\end{equation}\\

\noindent If PhENN were doing something trivial, e.g. edge sharpening, it would be failing to catch the phase shifts due to (\ref{eq:discon-frequency}). Therefore, the star pattern test also provides a way to test whether the physical model has been correctly incorporated. In Section \ref{subsec:star} we show experimental results verifying that indeed ImageNet-trained PhENN incorporates the physics whereas MNIST-trained PhENN does not.

\section{Experimental results} \label{sec:exp}

\subsection{Optical Apparatus}\label{subsec:apparatus}
\begin{figure}[ht!]
    \centering
    \setcaptionmargin{0.5in}
    \includegraphics[width = \textwidth]{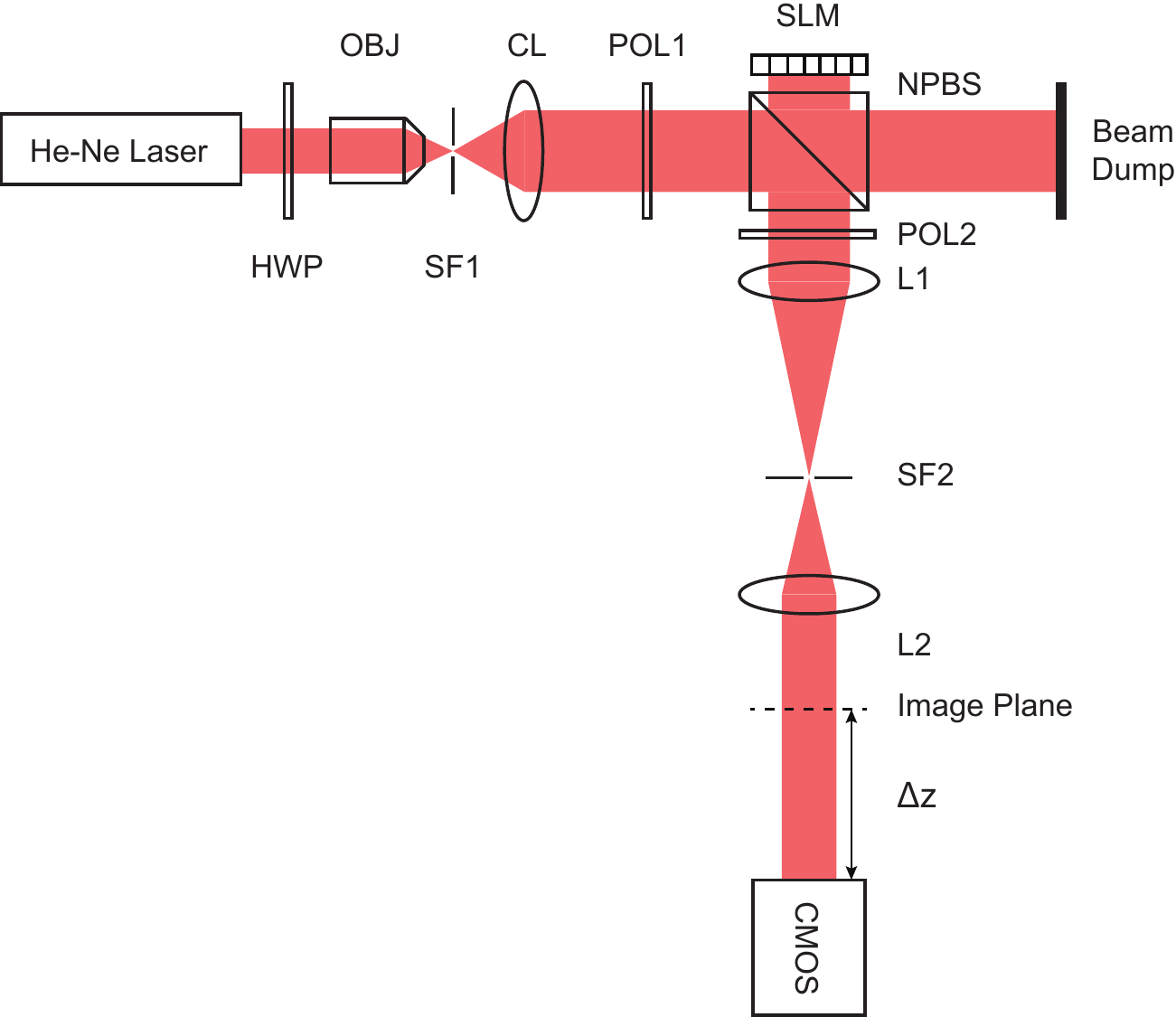}
    \caption{Optical apparatus. HWP: Half-wave plate, OBJ: objective lens, SF: spatial filter, CL: collimating lens, POL: linear polarizer, SLM: spatial light modulator, NPBS: non-polarizing beamsplitter, L: lens.}
    \label{fig:optical_apparatus}
\end{figure}

\begin{figure}[ht!]
    \centering
    \includegraphics[width=\textwidth,height=10cm]{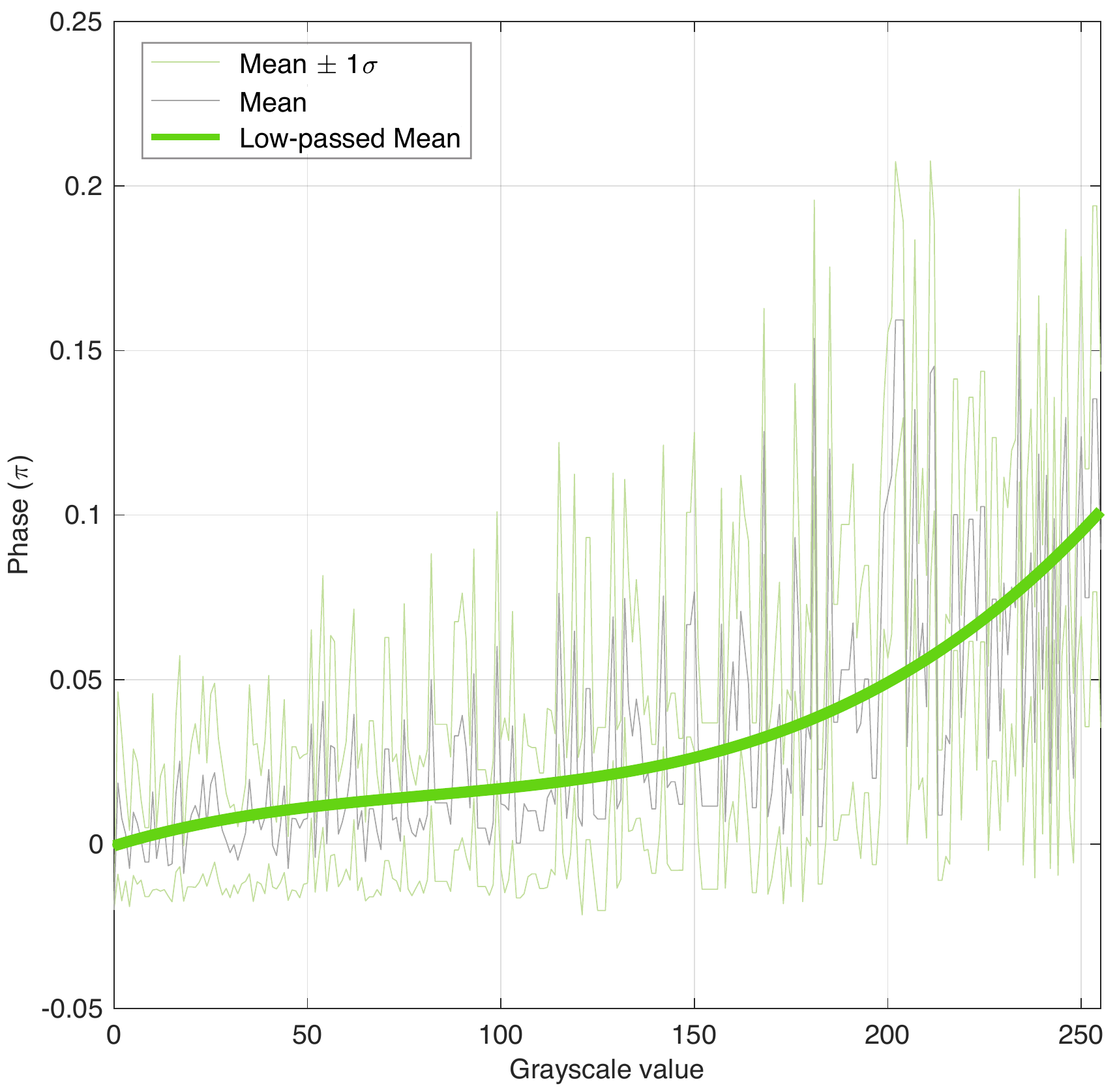}
    \caption{Phase modulation vs. 8-bit grayscale value for the experiments.}
    \label{fig:calibration}
\end{figure}

\noindent The optical configuration of the experiments is shown in Fig.~\ref{fig:optical_apparatus}. Polarization angles of two linear polarizers ($\text{POL}1$ and $\text{POL}2$) were carefully chosen to minimize the maximum phase modulation depth of $\text{SLM}$ (Holoeye LC-R 720) down to $\sim 0.1\pi$ (see Fig.~\ref{fig:calibration}) using Michelson-Young interferometer. Also, a $4f$ telescope of two lens $\left(f = 100\ \text{mm}\ \text{and}\ f = 60\ \text{mm}\right)$ was used to transfer the image plane from SLM matching two different pixel pitches of the SLM and CMOS camera, followed by a defocus of $ z = 150\ \text{mm}$ was given for capturing diffraction patterns with the camera. Each experimental diffraction pattern was iteratively registered with the simulated one by applying affine transformation on the experimental one to the direction of maximizing NMI (Normalized Mutual Information) between the two using Nelder-Mead method \cite{automaticregistration1999, neldermead1965}. More details on the pre-processing process is available in Appendix D. 

\subsection{Comparisons of the cross-domain generalization performance} \label{subsec:exp-generalization}
\begin{figure*}[ht!]
    \centering
    \setcaptionmargin{0.5in}
    \includegraphics[width=\textwidth]{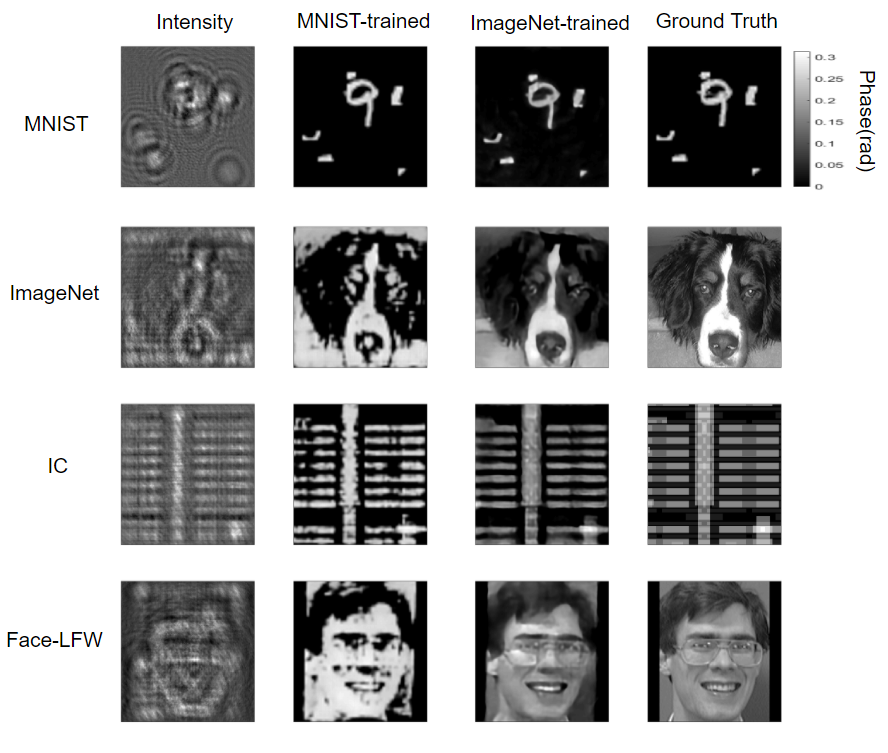}
    \caption{Cross-domain generalization performances of ImageNet-PhENN and MNIST-PhENN on experimental data. The defocus distance is $ z=150\text{mm}$.}
    \label{fig:cross-generalization-exp}
\end{figure*}

Cross-domain generalization performance based on experimental data ($z=150 \text{mm}$) for ImageNet-trained PhENN and MNIST-trained PhENN is shown in Fig.~\ref{fig:cross-generalization-exp} and quantitatively compared in Table~\ref{table:quantitative-cross-generalization-exp}, from which we make similar observations as those from the synthetic data: cross-domain generalization performance of ImageNet-trained PhENN is generally good; however, such performance of MNIST-trained PhENN is poor. Moreover, the estimated phase objects produced by MNIST-trained PhENN display sparse or binarized features, resembling MNIST examples, which suggests that during training, MNIST-trained PhENN was encouraged to memorize the MNIST examples, whereas ImageNet-trained PhENN, without strong regularization effect imposed by its training examples, was able to learn the actual physics model better. Additional examples are shown in Fig.~\ref{fig:cross-exp-Lz150-suppl} in Appendix C. 

\begin{table*}[hbt!]
\begin{center}
\begin{tabular}{|l||c|c||c|c|}
 \hline
& \multicolumn{2}{c||}{Average PCC $\pm$ std.dev}  & \multicolumn{2}{c|}{Average MAE $\pm$ std.dev} \\
 \hline
  & Train on ImageNet & Train on MNIST & Train on ImageNet & Train on MNIST \\
 \hline
Test on ImageNet & 0.908 $\pm$ 0.061 & 0.645 $\pm$ 0.233 & 0.008 $\pm$ 0.003 & 0.013 $\pm$ 0.006 \\
 \hline
Test on MNIST  & 0.956 $\pm$ 0.005 & 0.986 $\pm$ 0.007 & 0.005 $\pm$ 0.001 & 0.001 $\pm$ 0.0003 \\
 \hline 
Test on IC & 0.833 $\pm$ 0.035 & 0.736 $\pm$ 0.055 & 0.009 $\pm$ 0.001 & 0.012 $\pm$ 0.002 \\
\hline 
Test on Face-LFW & 0.951 $\pm$ 0.012 & 0.805 $\pm$ 0.056 & 0.006 $\pm$ 0.001 & 0.011 $\pm$ 0.002\\
\hline 
\end{tabular}
\caption{\textbf{Cross-domain generalization ability performance of PhENN trained with ImageNet and MNIST, respectively, for $ z=150 \text{ mm}$ (experimental data).}}
\label{table:quantitative-cross-generalization-exp}
\end{center}
\end{table*}

\subsection{Star-pattern experiment to demonstrate the learning of the propagation model}\label{subsec:star}

\begin{figure*}[ht!]
    \centering
    \setcaptionmargin{0.5in}
    \includegraphics[width=\textwidth]{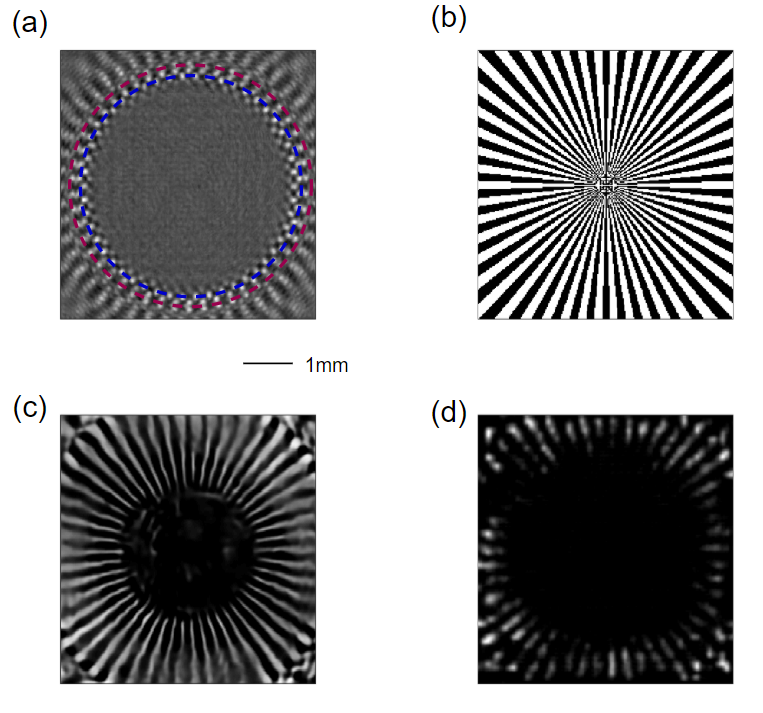}
    \caption{Reconstruction of the star pattern. (a). Intensity measurement at $z=150 \text{mm}$. (b). star pattern (weak) object. (c). reconstruction by ImageNet-trained PhENN. (d). reconstruction by MNIST-trained PhENN.}
    \label{fig:star-compare}
\end{figure*}

\noindent From Eq.~(\ref{eq:discon-frequency}), for our experimental parameters $ z=150\text{mm}, \lambda=633 \text{nm }, P=50$, the 2$^{\text{nd}}$ ($k=2$) and 3$^{\text{rd}}$ ($k=3$) discontinuity are at $0.0032 \mu\text{m}^{-1}$ and $0.0040 \mu \text{m}^{-1}$, respectively. From the measurement in Fig.~\ref{fig:star-compare}a, we see the two discontinuities are located at $r_2\approx 2.44 \text{mm}$ (red) and $r_3\approx 2.16 \text{mm}$ (blue), respectively, corresponding to spatial frequencies, $0.0033 \mu\text{m}^{-1}$ and $0.0037 \mu\text{m}^{-1}$, matching well with theoretical values, and indicating that the weak object approximation holds well. After PhENN was trained with ImageNet, a dataset drastically different from the appearance of the star-pattern, such discontinuity was corrected perfectly in the reconstruction (Fig.~\ref{fig:star-compare}c), indicating that ImageNet-trained PhENN has learned the underlying physics (or the WOTF) while the MNIST-PhENN apparently failed (Fig.~\ref{fig:star-compare}d). It is also noteworthy that there is still significant deficiency of high frequencies in the reconstruction of ImageNet-trained PhENN, which corroborates our observation earlier in Fig.~\ref{fig:WOTF-comparison-simulation} that even ImageNet-trained PhENN is not able to restore high frequencies very well. Using Learning-to-synthesize by DNN (LS-DNN) method \cite{deng2018learning,deng2020learning} has been proven very efficient to tackle this issue \cite{Lithesis}, but we choose not to pursue it as it would deviate from the main emphasis of this paper. \\

\section{Conclusions}
In this paper, we used PhENN for lensless phase imaging as the standard platform to address the important question of DNN generalization performance when  training cannot be performed in the intended class. This is motivated by the problem of insufficient training data, especially when such data need to be collected experimentally. We anticipate that this work will offer  practitioners a way to efficiently train their machine learning architectures, by choosing the publicly available standardized dataset without worrying about cross-domain generalization performance. \\

\noindent Our work is suggestive of certain interesting directions for future investigation. A particularly intriguing one is to refine the bound of the (cross-domain) generalization error by incorporating the distance metric of the empirical distributions between the training and the test set, along with other factors that have been considered currently in the literature (recall Section \ref{subsec:previous}). Moreover, though the chain of logic presented in the paper was centered on phase retrieval, it should be applicable to other domains of computational imaging subject to further study and verification.

\section*{Appendix A: More details of PhENN and the training specifics.}
\begin{figure}[ht!]
    \centering
    \setcaptionmargin{0.5in}
    \includegraphics[width=\textwidth]{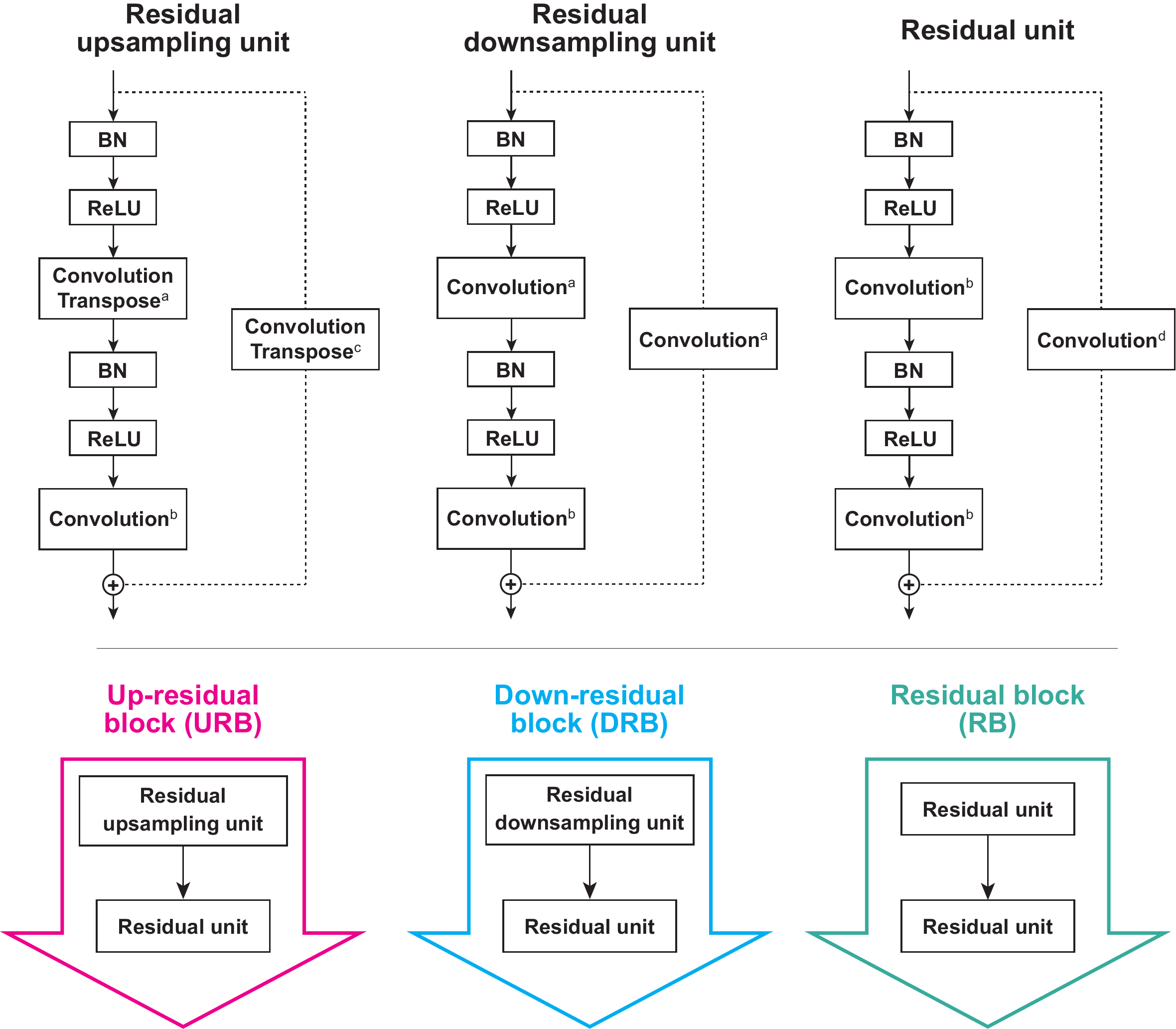}
    \caption{More detailed architecture of PhENN. Superscripts a - d denote different kernel size and strides, listed as follows: a) Kernel size: (3, 3), strides: (2, 2). b) Kernel size: (3, 3), strides: (1, 1). c) Kernel size: (2, 2), strides: (2, 2). d) Kernel size: (1, 1), strides: (1, 1).}
    \label{fig:PhENN-details}
\end{figure}

In Section \ref{subsec:phenn-intro}, we introduced the high-level architecture of PhENN (Fig.~\ref{fig:PhENN-generall}). In this Section, we provide in Fig.~\ref{fig:PhENN-details}  details of the layer-wise architecture of each functional block in PhENN. \\

\noindent In PhENN, major functional blocks include Up-residual blocks (URBs), Down-residual blocks (DRBs) and Residual blocks (RBs). All convolutional (Conv2D) and convolutional transpose (Conv2DTranspose) kernels are $3 \times 3$, except for the $2\times 2$ kernels in the side-branch Convolutional Transpose in Residual upsampling units and the $1 \times 1$ (1D convolution) kernels in the side-branch of Residual units. \\

\noindent The simulation is conducted on a Nvidia GTX1080 GPU using the open source machine learning Platform TensorFlow. The Adam optimizer \cite{inv:kingma2014adam}, with learning rate being 0.001 and exponential decay rate for the first and second moment estimates being 0.9 and 0.999 ($\beta_1$=0.9, $\beta_2$=0.999). The batch size is 5. The training of PhENN for 50 epochs, which is sufficient for PhENN to converge, takes about 2 hours.

 \section*{Appendix B: More results on synthetic data.}
\begin{figure}[hbt!]
    \centering
    \includegraphics[width=\textwidth]{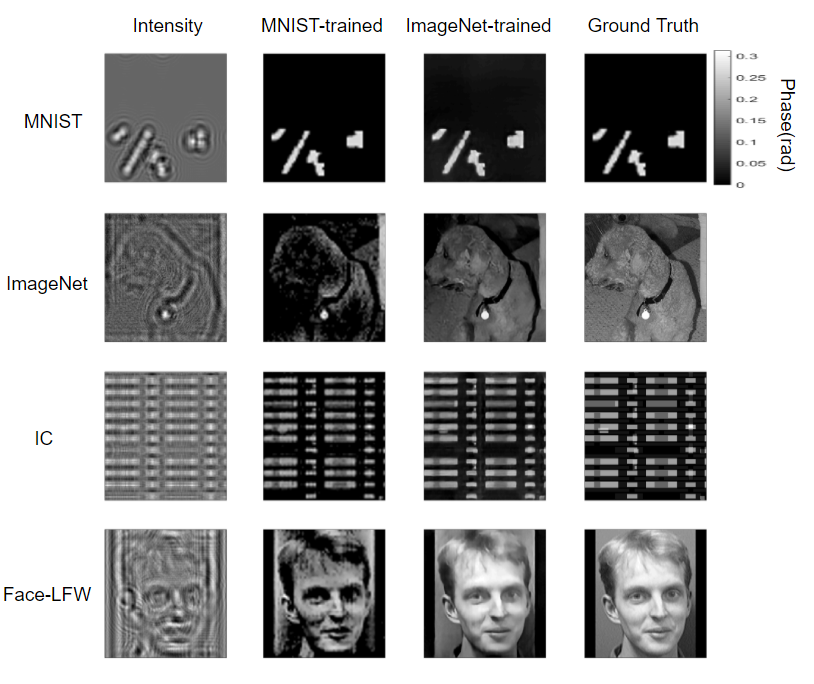}
    \caption{More examples of cross-domain generalization performance on synthetic data. The defocus distance $z=150\text{mm}$.}
    \label{fig:supple-cross-generalization-syn-150}
\end{figure}

In this section, we show reconstruction examples with synthetic data at $z=150\text{mm}$ in Fig.~\ref{fig:supple-cross-generalization-syn-150}, and the quantitative metrics in Table \ref{table:quantitative-syn-150}, where we see MNIST-trained PhENN produced reconstructions that are generally sparsified, as MNIST has imposed too strong regularization effect to the training and that gets passed onto the reconstructions. Superior cross-domain generalization performance of ImageNet-trained PhENN is again verified.

\begin{table*}[hbt!]
\begin{center}
\begin{tabular}{|l||c|c||c|c|}
 \hline
& \multicolumn{2}{c||}{Average PCC $\pm$ std.dev}  & \multicolumn{2}{c|}{Average MAE $\pm$ std.dev} \\
 \hline
  & Train on ImageNet & Train on MNIST & Train on ImageNet & Train on MNIST \\
 \hline
Test on ImageNet & 0.936 $\pm$ 0.043 & 0.586 $\pm$ 0.200 & 0.033 $\pm$ 0.023 & 0.055 $\pm$ 0.021 \\
 \hline
Test on MNIST  & 0.988 $\pm$ 0.005 & 0.9998 $\pm$ 0.0002 & 0.005 $\pm$ 0.001 & 0.001 $\pm$ 0.0004 \\
 \hline 
Test on IC & 0.911 $\pm$ 0.021 & 0.835 $\pm$ 0.060 & 0.021 $\pm$ 0.005 & 0.028 $\pm$ 0.006 \\
\hline 
Test on Face-LFW & 0.980 $\pm$ 0.011 & 0.747 $\pm$ 0.097 & 0.013 $\pm$ 0.003 & 0.042 $\pm$ 0.008\\
\hline 
\end{tabular}
\caption{\textbf{Cross-domain generalization ability performance of PhENN trained with ImageNet and MNIST, respectively, for $z=150 \text{ mm}$ (synthetic data).}}
\label{table:quantitative-syn-150}
\end{center}
\end{table*}

\section*{Appendix C: More results on experimental data.}
\begin{figure}
    \centering
    \setcaptionmargin{0.5in}
    \includegraphics[width=\textwidth]{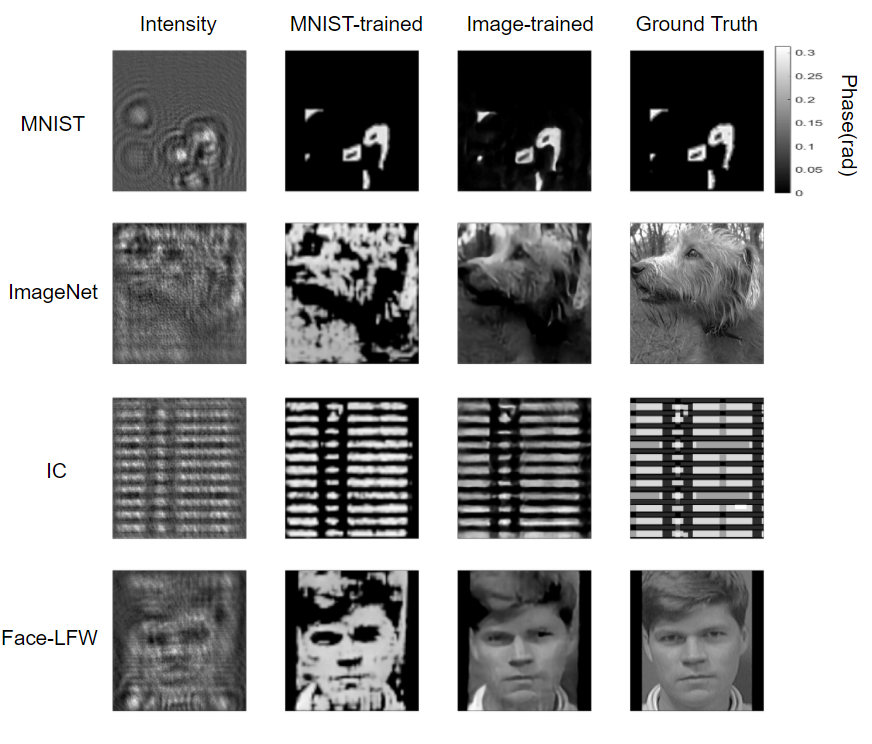}
    \caption{More examples of cross-domain generalization performance on experimental data. The defocus distance $z=150\text{mm}$.}
    \label{fig:cross-exp-Lz150-suppl}
\end{figure}

\noindent In Fig.~\ref{fig:cross-exp-Lz150-suppl}, we provide additional reconstructions of various classes of objects by ImageNet-trained PhENN and MNIST-trained PhENN ($z=150\text{mm}$), based on experimental data. Consistent with previous observations, we see significant distortions in the reconstructions of non-MNIST objects by MNIST-trained PhENN. However, no significant distortions can be seen from reconstructions of non-ImageNet objects produced by ImageNet-trained PhENN. 

\section*{Appendix D: Details on experimental data preprocessing.}

Two linear polarizers were used to achieve the maximum phase depth of the reflective SLM of $\sim 0.1\pi$, which however brings forth spurious effect on the shape of the exit beam. Thus, the training process with raw intensity measurements was preceded by the image registration on them.\\

\noindent Using the calibration curve between $8$-bit grayscale values and phase modulation depth in radian, input phase objects were simulated, followed by the generation of simulated intensity measurements without any deformation. Experimental measurements should match with the simulated ones if ideal. Image registration process finds an optimal affine transformation that brings an experimental measurement to its corresponding simulated one according to the optimization using Nelder-Mead method to the direction of minimizing the negative NMI (Normalized Mutual Information). Optimal affine transformation matrix was found for each dataset and applied accordingly.\\

\noindent Then, only center $256 \times 256$ pixels in each preprocessed intensity measurement were cropped to generate training, validation, and testing datasets, paired with labels or ground truth images. Thanks to the $4$f system in the optical apparatus that matches the pixel size of the reflective SLM with that of the CMOS camera, only the same dimension of pixels as that of each displayed image on the SLM is needed.\\

\section*{Funding Information}
Intelligence Advanced Research Projects Activity (IARPA), RAVEN Program (FA8650-17-C-9113); Singapore National Research Foundation, the SMART (Singapore-MIT Alliance for Research and Technology) program (015824). I. Kang was supported in part by the KFAS (Korea Foundation for Advanced Studies) scholarship.

\section*{Acknowledgments}

The authors thank Zhengyun Zhang for insightful discussions. 

\section*{Disclosures}

\noindent\textbf{Disclosures.} The authors declare no conflicts of interest.

\bibliography{main}

\begin{thebibliography}{10}
\newcommand{\enquote}[1]{``#1''}

\bibitem{inv:dong14-super-res}
{C. Dong}, {C. Loy}, {K. He}, and {X. Tang}, \enquote{Learning a deep
  convolutional neural network for image super-resolution,} in \emph{European
  Conference on Computer Vision (ECCV) / Lecture Notes on Computer Science Part
  IV,}  vol. 8692 (2014), pp. 184--199.

\bibitem{inv:perceptual-loss}
J.~Johnson, A.~Alahi, and {Li Fei-Fei}, \enquote{Perceptual losses for
  real-time style transfer and super-resolution,} in \emph{European Conference
  on Computer Vision (ECCV) / Lecture Notes on Computer Science,}  vol. 9906
  B.~Leide, J.~Matas, N.~Sebe, and M.~Welling, eds. (2016), pp. 694--711.

\bibitem{inv:ledig17}
C.~Ledig, L.~Theis, F.~Huczar, J.~Caballero, A.~Cunningham, A.~Acosta,
  A.~Aitken, A.~Tejani, J.~Totz, {Zehan Wang}, and {Wenshe Shi},
  \enquote{Photo-realistic single image super-resolution using a {Generative
  Adversarial Network},} in \emph{The IEEE Conference on Computer Vision and
  Pattern Recognition (CVPR),}  (2017), pp. 4681--4690.

\bibitem{inv:rivernson17-dlm}
Y.~Rivenson, Z.~Gorocs, H.~Gunaydin, {Yibo Zhang}, {Hongda Wang}, and A.~Ozcan,
  \enquote{Deep learning microscopy,} {\protect\JournalTitle{Optica}}
  \textbf{4}, 1437--1443 (2017).

\bibitem{inv:wang18-super-fluo}
{Hongda Wang}, Y.~Rivenson, {Zhensong Wei}, H.~Gunaydin, L.~Bentolila, and
  A.~Ozcan, \enquote{Deep learning achieves super-resolution in fluorescence
  microscopy,} bioRxiv, https://doi.org/10.1101/309641 (2018).

\bibitem{inv:nehme18-ML-STORM}
E.~Nehme, L.~E. Weiss, T.~Michaeli, and Y.~Shechtman, \enquote{{Deep-STORM:}
  super-resolution single-molecule microscopy by deep learning,}
  {\protect\JournalTitle{Optica}} \textbf{5}, 458--464 (2018).

\bibitem{deng2018learning}
M.~Deng, S.~Li, and G.~Barbastathis, \enquote{Learning to synthesize: splitting
  and recombining low and high spatial frequencies for image recovery,}
  {\protect\JournalTitle{arXiv preprint arXiv:1811.07945}}  (2018).

\bibitem{inv:sinha17-PhENN}
A.~Sinha, {Justin Lee}, {Shuai Li}, and G.~Barbastathis, \enquote{Lensless
  computational imaging through deep learning,} {\protect\JournalTitle{Optica}}
  \textbf{4}, 1117--1125 (2017).

\bibitem{inv:goy2018low}
A.~Goy, K.~Arthur, {Shuai Li}, and G.~Barbastathis, \enquote{Low photon count
  phase retrieval using deep learning,} {\protect\JournalTitle{Phys. Rev.
  Lett.}} \textbf{121}, 243902 (2018).

\bibitem{inv:WangH2018}
{H. Wang}, {M. Lyu}, and {G. Situ}, \enquote{eholonet: a learning-based
  point-to-point approach for in-line digital holographic reconstruction,}
  {\protect\JournalTitle{Opt. Express}} \textbf{26}, 22603--22614 (2018).

\bibitem{inv:pitkaaho17}
T.~Pitk{\"a}aho, A.~Manninen, and T.~J. Naughton, \enquote{Performance of
  autofocus capability of deep convolutional neural networks in digital
  holographic microscopy,} in \emph{Digital Holography and Three-Dimensional
  Imaging,}  (OSA, 2017), p. W2A.5.

\bibitem{inv:ozcan-dnn-extDOF}
{Y. Wu}, Y.~Rivenson, {Y. Zhang}, {Z.Wei}, H.~Gunaydin, {X. Lin}, and A.~Ozcan,
  \enquote{Extended depth-of-field in holographic image reconstruction using
  deep learning based auto-focusing and phase-recovery,}
  {\protect\JournalTitle{Optica}} \textbf{5}, 704--710 (2018).

\bibitem{ren2018learning}
Z.~Ren, Z.~Xu, and E.~Y. Lam, \enquote{Learning-based nonparametric
  autofocusing for digital holography,} {\protect\JournalTitle{Optica}}
  \textbf{5}, 337--344 (2018).

\bibitem{deng2020learning}
M.~Deng, S.~Li, A.~Goy, I.~Kang, and G.~Barbastathis, \enquote{Learning to
  synthesize: Robust phase retrieval at low photon counts,}
  {\protect\JournalTitle{Light: Science \& Applications}} \textbf{9}, 1--16
  (2020).

\bibitem{inv:deng2020probing}
M.~Deng, A.~Goy, S.~Li, K.~Arthur, and G.~Barbastathis, \enquote{Probing
  shallower: perceptual loss trained phase extraction neural network
  (plt-phenn) for artifact-free reconstruction at low photon budget,}
  {\protect\JournalTitle{Optics Express}} \textbf{28}, 2511--2535 (2020).

\bibitem{inv:horisaki16}
R.~Horisaki, R.~Takagi, and J.~Tanida, \enquote{Learning-based imaging through
  scattering media,} {\protect\JournalTitle{Opt. Express}} \textbf{24},
  13738--13743 (2016).

\bibitem{inv:IDiffNet}
{S. Li}, {M. Deng}, {J. Lee}, A.~Sinha, and G.~Barbastathis, \enquote{Imaging
  through glass diffusers using densely connected convolutional networks,}
  {\protect\JournalTitle{Optica}} \textbf{5}, 803--813 (2018).

\bibitem{li2018deep}
Y.~Li, Y.~Xue, and L.~Tian, \enquote{Deep speckle correlation: a deep learning
  approach toward scalable imaging through scattering media,}
  {\protect\JournalTitle{Optica}} \textbf{5}, 1181--1190 (2018).

\bibitem{kamilov2015learning}
U.~S. Kamilov, I.~N. Papadopoulos, M.~H. Shoreh, A.~Goy, C.~Vonesch, M.~Unser,
  and D.~Psaltis, \enquote{Learning approach to optical tomography,}
  {\protect\JournalTitle{Optica}} \textbf{2}, 517--522 (2015).

\bibitem{kamilov2016optical}
U.~S. Kamilov, I.~N. Papadopoulos, M.~H. Shoreh, A.~Goy, C.~Vonesch, M.~Unser,
  and D.~Psaltis, \enquote{Optical tomographic image reconstruction based on
  beam propagation and sparse regularization,} {\protect\JournalTitle{IEEE
  Transactions on Computational Imaging}} \textbf{2}, 59--70 (2016).

\bibitem{inv:goy19-3Dtomo}
A.~Goy, G.~Rughoobur, {Shuai Li}, K.~Arthur, A.~Akinwande, and G.~Barbastathis,
  \enquote{High-resolution limited-angle phase tomography of dense layered
  objects using deep neural networks,} {\protect\JournalTitle{Proc. Nat. Acad.
  Sci.}}  ((accepted) 2019).

\bibitem{inv:barbastathis19-review}
G.~Barbastathis, A.~Ozcan, and {Guohai Situ}, \enquote{On the use of deep
  learning for computational imaging,} {\protect\JournalTitle{Optica}}  (2019).

\bibitem{mccann2017convolutional}
M.~T. McCann, K.~H. Jin, and M.~Unser, \enquote{Convolutional neural networks
  for inverse problems in imaging: A review,} {\protect\JournalTitle{IEEE
  Signal Processing Magazine}} \textbf{34}, 85--95 (2017).

\bibitem{deng2009imagenet}
J.~Deng, W.~Dong, R.~Socher, L.-J. Li, K.~Li, and L.~Fei-Fei,
  \enquote{Imagenet: A large-scale hierarchical image database,} in \emph{2009
  IEEE conference on computer vision and pattern recognition,}  (Ieee, 2009),
  pp. 248--255.

\bibitem{nn:lecun2010mnist}
Y.~LeCun, C.~Cortes, and C.~J. Burges, \enquote{{MNIST} handwritten digit
  database,} {\protect\JournalTitle{AT\&T Labs [Online]. Available:
  http://yann. lecun. com/exdb/mnist}} \textbf{2} (2010).

\bibitem{tian2015quantitative}
L.~Tian and L.~Waller, \enquote{Quantitative differential phase contrast
  imaging in an led array microscope,} {\protect\JournalTitle{Optics express}}
  \textbf{23}, 11394--11403 (2015).

\bibitem{li2019analysis}
S.~Li, G.~Barbastathis, and A.~Goy, \enquote{Analysis of phase-extraction
  neural network (phenn) performance for lensless quantitative phase imaging,}
  in \emph{Quantitative Phase Imaging V,}  vol. 10887 (International Society
  for Optics and Photonics, 2019), p. 108870T.

\bibitem{inv:PhENN-spectral-premod}
{S. Li} and G.~Barbastathis, \enquote{Spectral pre-modulation of training
  examples enhances the spatial resolution of the phase extraction neural
  network ({PhENN}),} {\protect\JournalTitle{Opt. Express}} \textbf{26},
  29340--29352 (2018).

\bibitem{neyshabur2017exploring}
B.~Neyshabur, S.~Bhojanapalli, D.~McAllester, and N.~Srebro, \enquote{Exploring
  generalization in deep learning,} in \emph{Advances in Neural Information
  Processing Systems,}  (2017), pp. 5947--5956.

\bibitem{neyshabur2018towards}
B.~Neyshabur, Z.~Li, S.~Bhojanapalli, Y.~LeCun, and N.~Srebro, \enquote{Towards
  understanding the role of over-parametrization in generalization of neural
  networks,} {\protect\JournalTitle{arXiv preprint arXiv:1805.12076}}  (2018).

\bibitem{neyshabur2017pac}
B.~Neyshabur, S.~Bhojanapalli, and N.~Srebro, \enquote{A pac-bayesian approach
  to spectrally-normalized margin bounds for neural networks,}
  {\protect\JournalTitle{arXiv preprint arXiv:1707.09564}}  (2017).

\bibitem{zhang2016understanding}
C.~Zhang, S.~Bengio, M.~Hardt, B.~Recht, and O.~Vinyals, \enquote{Understanding
  deep learning requires rethinking generalization,}
  {\protect\JournalTitle{arXiv preprint arXiv:1611.03530}}  (2016).

\bibitem{advani2017high}
M.~S. Advani and A.~M. Saxe, \enquote{High-dimensional dynamics of
  generalization error in neural networks,} {\protect\JournalTitle{arXiv
  preprint arXiv:1710.03667}}  (2017).

\bibitem{xu2012robustness}
H.~Xu and S.~Mannor, \enquote{Robustness and generalization,}
  {\protect\JournalTitle{Machine learning}} \textbf{86}, 391--423 (2012).

\bibitem{jakubovitz2019generalization}
D.~Jakubovitz, R.~Giryes, and M.~R. Rodrigues, \enquote{Generalization error in
  deep learning,} in \emph{Compressed Sensing and Its Applications,}
  (Springer, 2019), pp. 153--193.

\bibitem{shannon1948mathematical}
C.~E. Shannon, \enquote{A mathematical theory of communication,}
  {\protect\JournalTitle{Bell system technical journal}} \textbf{27}, 379--423
  (1948).

\bibitem{cover2012elements}
T.~M. Cover and J.~A. Thomas, \emph{Elements of information theory} (John Wiley
  \& Sons, 2012).

\bibitem{automaticregistration1999}
G.~K. Matsopoulos, N.~A. Mouravliansky, K.~K. Delibasis, and K.~S. Nikita,
  \enquote{Automatic retinal image registration scheme using global
  optimization techniques,} {\protect\JournalTitle{IEEE Transactions on
  information technology in biomedicine}} \textbf{3}, 47--60 (1999).

\bibitem{neldermead1965}
J.~A. Nelder and R.~Mead, \enquote{A simplex method for function minimization,}
  {\protect\JournalTitle{The computer journal}} \textbf{7}, 308--313 (1965).

\bibitem{Lithesis}
S.~Li, \enquote{Computational imaging through deep learning,} Ph.D. thesis, MIT
  (2019).

\bibitem{inv:kingma2014adam}
D.~P. Kingma and {J. Lei Ba}, \enquote{Adam: A method for stochastic
  optimization,} in \emph{International Conference on Learning Representations
  (ICLR),}  (2015).

\end{thebibliography}



\end{document}